# Unravelling the intertwined atomic and bulk nature of localised excitons by attosecond spectroscopy

Matteo Lucchini[1,2*], Shunsuke A. Sato[3,4], Giacinto D. Lucarelli[1,2], Bruno Moio[1], Giacomo Inzani[1], Rocío Borrego-Varillas[2], Fabio Frassetto[5], Luca Poletto[5], Hannes Hübener[4], Umberto De Giovannini[4,6], Angel Rubio[4,6], Mauro Nisoli[1,2]

[1]Department of Physics, Politecnico di Milano, 20133 Milano, Italy
[2]Institute for Photonics and Nanotechnologies, IFN-CNR, 20133 Milano, Italy
[3]Center for Computational Sciences, University of Tsukuba, Tsukuba 305-8577, Japan
[4]Max Planck Institute for the Structure and Dynamics of Matter, 22761 Hamburg, Germany
[5]Institute for Photonics and Nanotechnologies, IFN-CNR, 35131 Padova, Italy
[6] Nano-Bio Spectroscopy Group, Universidad del País Vasco, 20018 San Sebastian, Spain

*To whom correspondence should be addressed; E-mail: matteo.lucchini@polimi.it.

**The electro-optical properties of most semiconductors and insulators of technological interest are dominated by the presence of electron-hole quasiparticles called excitons[1,2]. The manipulation of these hydrogen-like quasi-particles in dielectrics, has received great interest[3] under the name *excitonics*[4] that is expected to be of great potential for a variety of applications, including optoelectronics and photonics[5–7]. A crucial step for such exploitation of excitons in advanced technological applications is a detailed understanding of their dynamical nature. However, the ultrafast processes unfolding on few-femtosecond and attosecond time scales, of primary relevance in view of the desired extension of electronic devices towards the petahertz regime, remain largely unexplored.**

Here we apply attosecond transient reflection spectroscopy[8] in a sequential two-foci geometry[9] and observe sub-femtosecond dynamics of a core-level exciton in bulk MgF$_2$ single crystals. With our unique setup, we can access absolute phase delays which allow for an unambiguous comparison with theoretical calculations based on the Wannier-Mott model[10,11]. Our results show that excitons surprisingly exhibit a dual atomic- and solid-like character which manifests itself on different time scales. While the former is responsible for a femtosecond optical Stark effect[12,13], the latter dominates the attosecond excitonic response and originates by the interaction with the crystal. Further investigation of the role of exciton localization proves that the bulk character persists also for strongly localised quasi-particles and allows us to envision a new route to control exciton dynamics in the close-to-petahertz regime.

The quest for new devices capable of surpassing the current technological limits[14] has pushed the scientific community to explore solutions beyond classical electronics as done in excitonics[4], spintronics and valleytronics[15]. Therefore, studying the dynamics of excitons in solids becomes a priority task not only to widen our knowledge of fundamental solid-state dynamical phenomena, but also to explore the ultimate limits of these novel technologies[16,17]. While the development of attosecond spectroscopy[18], has proven the possibility to study sub-femtosecond electron dynamics in solids[19], shedding new light onto strong field phenomena and light-carrier manipulation [20–22], a clear observation of attosecond exciton dynamics was missing. Besides more conventional femtosecond techniques[23], attosecond transient absorption and reflectivity spectroscopy have been employed to study the ultrafast decay processes (few-femtoseconds) of core-excitons[24,25], but failed in recording the sub-cycle dynamics unfolding during light-matter interaction. Here we used attosecond transient reflection spectroscopy (ATRS)[8] to study attosecond dynamics of a core-level exciton in bulk $MgF_2$ single crystals, characterized by a binding energy similar to the interlayer excitons of two-dimensional materials[17]. Thanks to the employment of simultaneous and independent calibration experiments, we achieved direct comparison with theoretical simulations, which in turn allowed us to make a clear link between the observed transient features of the system optical response and the nanometric and attosecond motion of the excitons. Our results thus provide a novel description of the ultrafast exciton-crystal interaction on sub-femtosecond time scales and move an important step forward attosecond excitonics.

Figure 1a shows a schematic picture of the experimental setup characterized by a sequential two-foci geometry, used to perform simultaneous attosecond photoelectron and ATRS measurements[9] (see Methods). The static reflectivity for a $MgF_2$ (001) crystal, $R_0$, close to the Mg $L_{2,3}$ edge is probed by an extreme-ultraviolet (XUV) attosecond pulse (Fig. 1b). In this energy region $R_0$ is characterized by a peak (labelled with A), which has been attributed to the formation of excitons after excitation of a localized $Mg^{2+}$ $2p$ core electron[26] (Fig. 1c). The weaker satellite peak at about 54 eV (A' in Fig. 1b) originates instead from the spin-orbit splitting of the $Mg^{2+}$ $2p$ core state[27]. In our experiment we use a few-femtosecond and intense IR pulse to drive the crystal out of equilibrium (Fig. 1c), initiating ultrafast exciton dynamics later probed by monitoring the sample differential reflectivity ΔR/R in the XUV range, defined as the difference between pumped and unpumped reflectivity, divided by the latter: [$R_{IR}$ - $R_0$] /$R_0$. The results as a function of photon energy $E$ and pump-probe delay $t$, are shown in Fig. 1d. At small values of $t$, we observe rich transient features which can be decomposed in a slower and a faster component. While the former unfolds on a few-femtosecond time scale and is mainly located around the excitonic features A and A', the latter oscillates at twice the IR frequency and extends over the full energy range under consideration, becoming more evident in the conduction band (CB) region. The upper panel in Fig. 1d shows the square of the IR vector potential, $A^2_{IR}$, as retrieved from the simultaneous streaking trace (see supplementary material). Knowing the associated pump field time evolution, we calibrated the pump-probe delay axis in order to have the zero delay coinciding with the maximum of the IR electric field squared (equivalently, a zero of $A^2_{IR}$). This allowed us to set an absolute reference for our measurements and study the precise timing of the system dynamics.

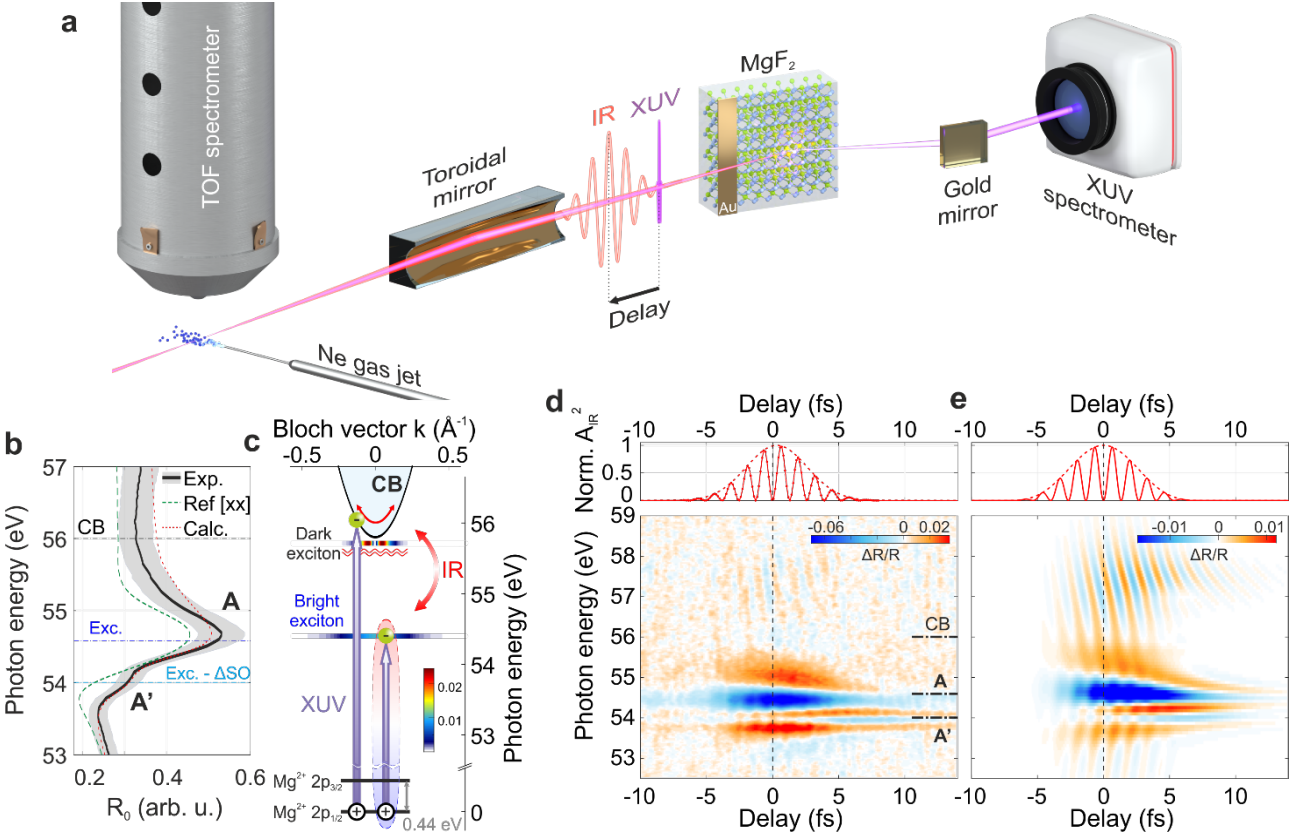

*Fig. 1 | MgF$_2$ core-exciton optical response*. **a**, Scheme of double-foci the experimental setup composed by the Ne gas target to perform an attosecond streaking experiment, the MgF2 crystal and the XUV spectrometer. **b**, Experimental (black solid) and calculated (red dotted) MgF$_2$ reflectivity for an incidence angle of 73.5°. The grey shaded area represents twice the standard deviation over repeated experimental measurements while the green dashed curve is the reflectivity extracted from (22). The horizontal dash-dotted lines mark the vertical transitions from the Mg$^{2+}$2p core level to the conduction band (CB) and the bright exciton (21), accounting for the spin-orbit splitting of the Mg$^{2+}$ 2p state. **c**, MgF$_2$ band structure around its CB. While XUV photons can promote electrons from the Mg-2p state directly into the CB, or to an electron-hole pair state (exciton), the IR field perturbs the crystal, causing phenomena like intra-band motion or dressing of dark excitonic states. The false colours represent the density of states for the excitons. Experimental, **d**, and calculated, **e**, transient reflection spectrograms (main panels) as a function of the delay between the XUV and IR pulses, together with the square of the IR vector potential, $A^2_{IR}$, as extracted from the simultaneous streaking measurement (upper panels). The delay zero (vertical dashed line) has been chosen in order to coincide with a maximum of the IR electric field squared.

To reach a complete understanding, we calculated the quantum dynamics with a Wannier-Mott (WM) exciton model[10,11] (see Methods) and chose the same reference for the delay zero, such that we could directly compare experimental and theoretical results. The calculated ΔR/R reported in Fig. 1e accurately reproduces the experimental data of Fig. 1d, suggesting dynamics which go beyond the optical Stark effect.

The slow component of the experimental ΔR/R (Fig. 2a) is characterized by a series of positive (red) and negative (blue) features, which develop around delay zero and fully disappear within 10-15 fs. These features originate mainly from IR-induced optical Stark effect of the excitonic transitions, from which it is possible to evaluate the exciton decay time[24] and its coupling with phonons[25]. For the main excitonic transition A, we found the Auger decay term[28] to dominate, giving a very short decay time of 2.35 ± 0.3 fs (see Methods), in agreement with what observed for other insulators like $SiO_2$ and $MgO_2$. While a detailed description of these features is beyond the scope of this work, it is important to stress that they can be explained considering the exciton to behave like an atom. Since excitons in $MgF_2$ have an atomic-localised nature, solid-like phenomena are not observed, and the dominant effect of the IR pump is a time-dependent blue-shift.

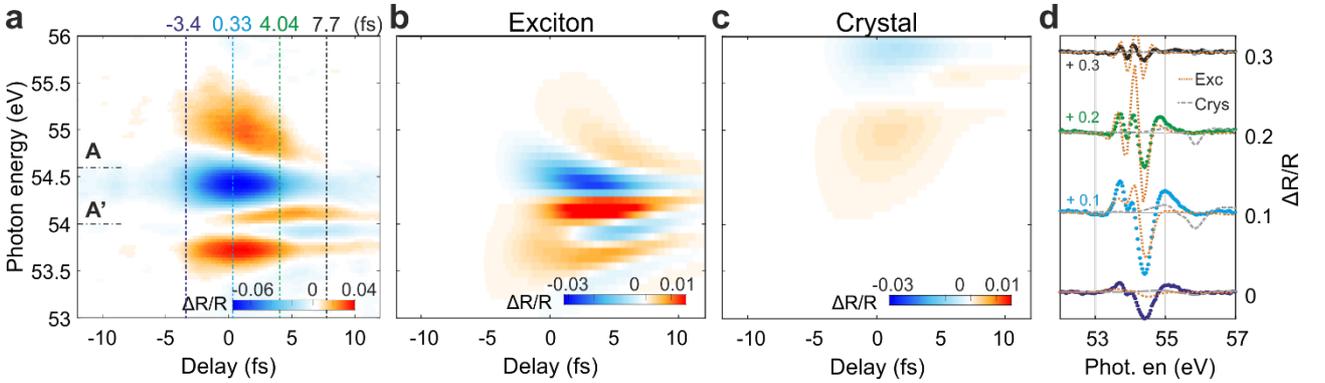

*Fig. 2 | **Real-time observation of femtosecond core-exciton dynamics**. **a**, Slow component of the experimental ΔR/R reported in Fig. 1d. **b, c**, Calculated slow component of ΔR/R considering solely the two excitonic states or only the crystal CB, respectively. **d**, Measured (dots) and calculated (orange dotted and gray dashed curves) ΔR/R profiles for four representative delays marked by the dash-dotted vertical lines in a.*

Further insight can be obtained by comparison with our simulations. Figures 2b, 2c show the calculated slow component of ΔR/R considering only the bright and dark excitonic states in Fig. 1c ("pure excitonic" contribution) or only the crystal CB ("pure crystal" contribution) (see Methods). We find that "pure exciton" calculations exhaustively reproduce the experimental dispersive profiles (Fig. 2d), while the crystal response does not. This further supports the attribution of the origin of the slow component of ΔR/R to the optical Stark effect and reflects the expected atomic-like character of the exciton quasi-particle.

While this atomic character is typical for an exciton characterized by a binding energy of 1.4 eV [29], the analysis of the sub-fs dynamics reveals an unexpected result. The fast component of the transient reflectivity spectrogram is reported in Fig. 3a, showing clear oscillations at twice the IR frequency, which are fully reproduced by our simulations (Fig. 3b). These oscillations have a V-shaped dispersion which resembles what is found for the dynamical Franz-Keldysh effect (DFKE) high into the CB of diamond[30,31]. While this suggests a clear link with intra-band motion of virtual charges, we note that the present case differs significantly as the V-shaped structure in MgF$_2$ is located across the CB bottom and centred on the excitonic transition A, where the presence of real carriers can modify the system response[21,32].

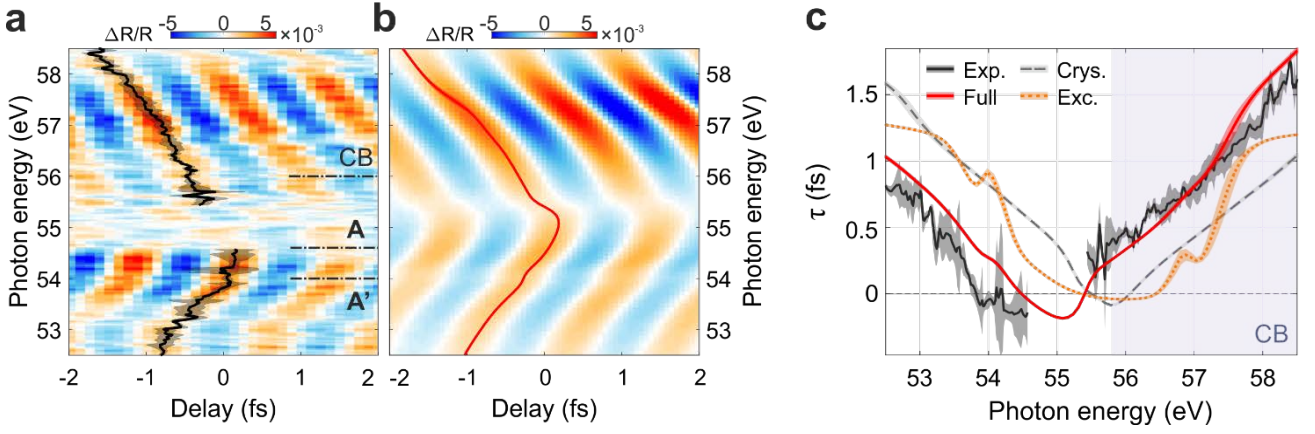

*Fig. 3 | **Core-exciton dynamics on an attosecond time scale**. **a**, Fast component of the experimental ΔR/R. The phase delay, τ, between the oscillations and E$_{IR}$² is displayed in black. The solid curve represents the mean over four independent measurements and the shaded area twice its standard deviation. **b**, Same as a but for the full calculation results. **c**, Comparison between the experimental (black solid) and calculated τ, considering the full system response (red solid), or addressing separately the crystal (grey dashed) and excitonic (orange dotted) contributions.*

Thanks to our unique experiment we can make a further step towards a complete comprehension of such a rich exciton dynamics and study the phase delay τ between the oscillations in the transient signal and the square of the IR electric field E$^2_{IR}$ (see Methods). Figure 3c presents the experimental τ obtained as a weighted average over four independent measurements (black solid curve) compared with the calculated phase delay in case of full model (red solid curve), pure exciton (orange dotted curve) or pure crystal response (dashed grey curve). In all cases, the shaded area represents twice the standard deviation originating from the measurement error or from the uncertainty of the phase extraction method. The full model accurately reproduces the experiment both on a qualitative and quantitative level. However, in contrast to what we observed for the slow component of ΔR/R, now the pure excitonic contribution alone fails in capturing the measured τ, even qualitatively. The system response is instead closer to the bare crystal bulk-case, showing the same shape, but shifted towards lower photon energies. This strongly indicates that the fast component of the differential reflectivity is dominated by the solid nature of the exciton, despite its localised character and in stark contrast to its slower atomic like response. Thus, by addressing separately the fast (attosecond) and slow (femtosecond) components of the optical response with our time-resolved attosecond measurements, we are able to disentangle the optical Stark effect (atomic-like) and the DFKE (solid-like) in excitons, which were previously thought to compete in time-averaged measurements[33].

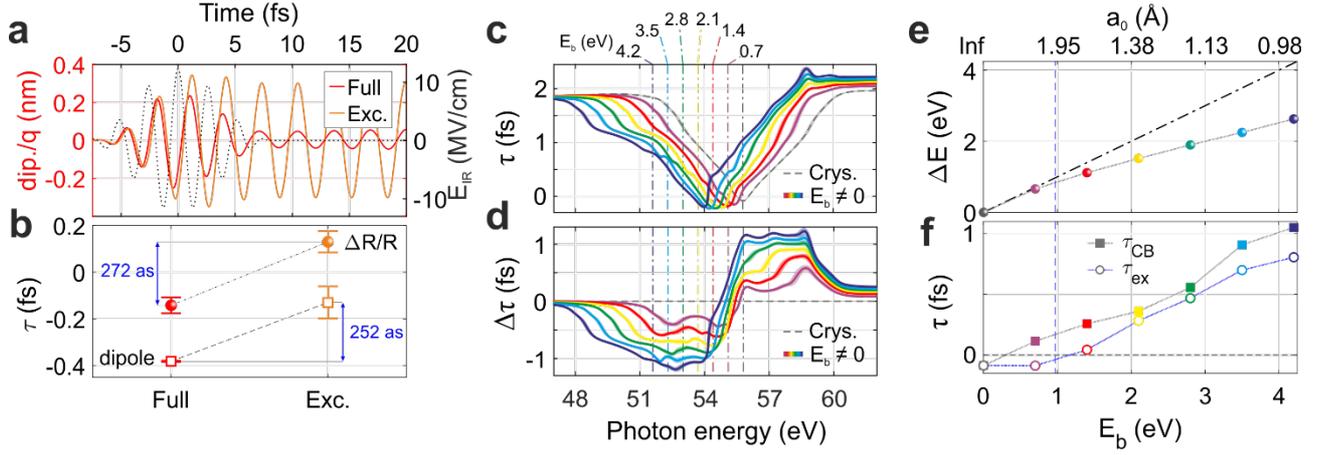

Fig. 4 | **Exciton real space dynamics and role of localization. a,** Exciton dipole oscillations in real time computed with the full (red) or pure exciton (orange) model. The black-dotted curve represents the IR electric field. **b,** Comparison of the phase delay extracted from ΔR/R (circles) or from the dipole moment (squares) for the full calculations (red) or considering only the excitonic response (orange). **c,** Behaviour of the calculated phase delay τ as a function of the exciton binding energy $E_b$. The vertical dash-dotted lines mark the exciton position. **d,** Difference between the phase delay of the full system and the pure crystal response, Δτ, for the data reported in f. **e,** Energy shift ΔE which minimizes Δτ as a function of $E_b$ (Bohr radius $a_0$). The vertical blue dashed line marks the minimum inter-nuclear distance in the MgF$_2$ crystal unit cell equal to 1.98 Å. **f,** Value of the phase delay τ evaluated at the exciton vertical transition ($τ_{ex}$, open circles), or at the bottom of the CB ($τ_{CB}$, full squares) for the different values of $E_b$ ($a_0$) considered.

Whether the observed transient features in the optical response correspond to an actual movement of the exciton on attosecond and nanometric scales is an open question. To tackle this, we calculated the excitonic dipole in real time and found that it oscillates almost at the same frequency of the IR field during interaction. The results are shown in Fig. 4a for the case of the full system (red curve) or considering only the quasi-particle (orange curve). By evaluating the phase delay for the dipole with respect to the IR field (black dotted curve), in the same fashion as done for ΔR/R, we found the oscillations of the pure exciton dipole to be delayed by 252 ± 69 as with respect to the full system response (square marks in Fig. 4b). This value is in agreement with the difference of 272 ± 57 as observed between the phase delay τ of Fig. 3c, evaluated at the energy of the excitonic transition and calculated with the full model or considering only the excitonic contribution (full circles in Fig. 4b). Therefore, our findings suggest a strong correlation between the sub-nm exciton motion in real space and the sub-fs transient features observed in the differential reflectivity. The few-cycle IR pulse dresses the crystal CB inducing intra-band motion (DFKE). In turn, this alters the exciton dynamical properties causing the quasi-particle to oscillate almost in phase with $A_{IR}$. This not only proves that localized excitons can show properties beyond atomic model, widening our comprehension of the ultrafast solid-state physics, but also opens the possibility to investigate separately different competing physical mechanisms, ultimately allowing for a full optimization of light-matter interaction in future devices.

Even if we proved that exciton-crystal interaction can transfer bulk properties to atomic-like excitons on attosecond time scale, the question remains whether this dual nature is affected by the localization degree of the quasi-particle. In order to further investigate the role of real space dynamics and exciton localization we computed the reflectivity phase delay τ for different exciton binding energies $E_b$ or, equivalently, Bohr radii $a_0$, which provides a reference for the actual size of excitonic devices[6,34]. With increasing $E_b$ and thus the degree of localization, τ preserves its V shape which centre appears to move towards lower photon energies (Fig. 4c), resulting in an overall bigger phase delay difference, Δτ, between the bare crystal and full system responses (Fig. 4d). The energy shift ΔE which minimizes Δτ is not exactly equal to $E_b$ (Fig. 4e), indicating that τ does not simply experience a rigid shift following the exciton transition. On the one hand this further confirms that the central role of exciton-crystal interaction, on the other hand it opens a new way to sculpt and control the excitonic response in the petahertz regime. While the optical response of delocalized electron-hole pairs will react almost in phase with the pump electric field ($τ_{ex}$ in Fig. 4f), more localized excitons will have a slower response, almost out of phase. Due to the exciton-crystal interaction, the CB response will also be significantly affected (see $τ_{CB}$ in Fig. 4f). Since the exciton binding energy can be continuously tuned around its natural value by an external field, modifying the dielectric screening or inducing a strain[35–37], one can control the phase delay between the quasi-particle and an optical field on an attosecond level. In particular, for values of $E_b ≈ 1$ eV where $a_0$ becomes comparable with the minimum inter-nuclear distance in the $MgF_2$ unit cell, it is possible to control the attosecond timing between the exciton and the CB signal, realizing a condition when the first is advanced with respect to the IR field while the second is delayed. As those are the typical binding energies of interlayer excitons in Van der Waals heterostructures[17], our findings point to a possible route for the realization of new devices where a control over the degree of Coulomb screening[38] can be used to tune the system response timing with attosecond resolution.

We investigated ultrafast core-exciton dynamics around the $L_{2,3}$ edge in $MgF_2$ single crystal with ATRS. Simultaneous calibration measurements allowed us to perform a direct and unambiguous comparison with theoretical results, addressing the dual nature of the excitonic quasi-particle from a new perspective. In particular, we found that while the exciton dynamics unfolding on the first few femtoseconds originate from the optical Stark effect and can be understood invoking just the atomic character of the quasi-particle, the physical processes happening on an attosecond time scale during light-matter interaction (i.e. intra-band motion and DFKE) are typical of the condensed state of matter. Moreover, our theoretical simulations and analysis revealed that the atom-solid duality is general and exists also for strongly bounded excitons where the absolute timing of the system optical response can be controlled on attosecond time scale by tuning the exciton binding energy. These findings set a new lever to the development of innovative optical devices for petahertz optoelectronics mediated by excitonics.

**Methods**

**Experimental setup:**

The setup used for the experiment reported in the main manuscript is described in detail in Ref. [9]. Single attosecond pulses (SAPs), centred around 42-eV photon energy, and 5-fs infrared (IR) pulses (central wavelength 750 nm) are first focused onto a Ne gas target. The XUV pulses have a time duration of about 250 as while the IR peak intensity is set between $10^{12}$ and $10^{13}$ W/cm². A time-of-flight (TOF) spectrometer records the photoelectron spectra as a function of the delay, $t$, between the IR and XUV pulses to perform an attosecond streaking experiment[39]. This allows us to retrieve the temporal characteristics of the two pulses and to obtain a precise calibration of the relative delay $t$ by extracting the exact shape of the IR vector potential. A gold-plated toroidal mirror then focuses both beams onto the MgF$_2$ crystal, where a thin gold layer deposited on a portion of the sample is used to calibrate the incident XUV photon flux and extract the energy-dependent sample reflectivity, $R_0(E)$. For more details, see the related sections in the Supplementary Information.

**Data analysis:**

To study the different mechanisms underlying the transient features observed in $\Delta R/R$, we decompose the pump-probe spectrogram in a slow and a fast component. To extract the slow component, we apply a low-pass frequency filter to the reflectivity spectrogram which is constant for frequencies below a cut-off frequency fc and decays with a supergaussian a profile $e^{\left(\frac{f-f_c}{2\sigma_f}\right)^n}$ with coefficient n = 16 and width $\sigma_f = 0.01$ PHz. Since the fastest feature observed oscillates at twice the IR frequency $2f_{IR} \approx 0.75$ PHz, we decided to set fc to $1.5f_{IR} = 0.5621$ PHz. Once the slow component has been extracted, the fast component of $\Delta R/R$ is simply obtained by subtracting the slow component from the total spectrogram. In the case of the experimental data, a high-frequency filter centred at $5f_{IR} = 1.8737$ PHz is used to remove the fast noise from the data prior to slow and fast decomposition.

As discussed in the main manuscript, the femtosecond transient features of $\Delta R/R$ originate from the optical Stark effect (OSE) induced by the IR electric field. To extract the Stark shift $\varepsilon$ from the experimental data, at each delay $t$, we fitted the sample reflectivity at the presence of the IR pump, $R_{IR}(E, t)$ with six Gaussian bells. Two Gaussians describe the background. Their parameters are derived from the static reflectivity $R_0(E)$. The other four Gaussians are used to fit the bright and dark exciton features, doubled because of the Mg$^{2+}$ 2p spin-orbit splitting. As observed for MgO[25], the dark excitonic state is responsible for an increase of $R_{IR}(E, t)$ around $t = 0$ fs, which appears next to the bright excitonic peak, on the low energy side, thus overlapping with the bright exciton signal which originates from $2p_{3/2}$ state. Due to the energy overlap, it is not possible to fit accurately the contribution of the $2p_{1/2}$-dark state transition as well as all the transitions involving the $2p_{3/2}$ state. Therefore, we can obtain a reliable estimation of ε(t) only for the bright-exciton - $2p_{1/2}$ transition which is found to follow the delay-dependent energy position of the maximum of $\Delta R/R$ around the A feature. The excitonic dipole $d(t)$ is obtained by deconvoluting the delay-dependent Stark shift ε(t) with the envelope of the IR electric field[25], directly extracted from the simultaneous streaking trace. The Auger decay rate γ and the phonon coupling φ can then be evaluated by modelling the excitonic dipole with the function $d(t) \sim e^{-\gamma t} e^{\phi(t)}$ [24,25].

The absolute phase delay between the fast transient feature of $\Delta R/R$ and the IR electric field is evaluated following the approach reported in [21,31]. First the IR vector potential is extracted from the simultaneous streaking trace by means of a 2D fitting procedure based on the analytical model reported in [40]. Then the phase difference between the transient features in the differential reflectivity and $E_{IR}^2$ is directly evaluated by multiplying the energy dependent Fourier transform of the first with the complex conjugate of the Fourier transform of the latter. The product thus constructed automatically peaks at the common frequency between the signals and has a phase equal to their phase difference $\Delta\varphi$. As $\Delta\varphi$ could be frequency-dependent, we evaluate the average value around $2f_{IR}$ using the local intensity of the Fourier transform product as weight. The standard deviation is obtained calculating the second momentum of this distribution. Finally, the phase delay $\tau$ is given by the ration between $\Delta\varphi$ and the beating frequency $2f_{IR}$. The main results reported in Fig. 3c represent average over 4 independent transient reflection measurements conducted under similar conditions and weighted by the inverse of their individual experimental uncertainty. The final error accounts both for the mean measurement error and for the statistical deviation between the independent measurements. For further details, see the Supplementary Information. We note that $\tau$ has the opposite sign of the pump-probe delay modulations (compare the black curves in Fig. 3a and Fig. 3c). This originates from the fact that a positive pump-probe delay $t$ means that the IR pulse is coming later (IR behaving as a probe), but for the sake of an easier interpretation, we chose a positive $\tau$ to mean that the system as a delayed response with respect to the IR field in real time (IR behaving as a pump).

**Theoretical model:**

To investigate the microscopic mechanism of the experimental observation, we describe the dynamical system under the intense IR pulse and the weak XUV pulse based on the following ansatz:

$$|\Psi(t)\rangle = |\Phi_{GS}\rangle + \sum_k c_k(t)\, \hat{a}^\dagger_{c,k+eA_{IR}(t)/\hbar c}\, \hat{a}_{v,k+eA_{IR}(t)/\hbar c}|\Phi_{GS}\rangle, \qquad (1)$$

where $|\Psi(t)\rangle$ is the wavefunction of the dynamical system, $|\Phi_{GS}\rangle$ is the ground state wavefunction of the matter, $\hat{a}^\dagger_{c,k}$ ($\hat{a}_{v,k}$) is a creation (annihilation) operator for the conduction (valence) state at the Bloch wavenumber, $k$, and $c_k(t)$ is an expansion coefficient. Here, the vector potential of the IR field is denoted as $A_{IR}(t)$. Note that the ansatz in Eq. (1) is a linear combination of single-particle single-hole states and is in line with the Tamm-Dancoff approximation. Treating the excitation from the ground state to electron-hole states by the XUV field perturbatively, the equation of motion for the coefficient $c_k(t)$ is given by

$$i\hbar \dot{c}_k(t) = \left[\sum_{k'} H_{kk'}(t) c_{k'}(t)\right] + E_{XUV}(t) \cdot D_{k+eA_{IR}(t)/\hbar c}, \qquad (2)$$

where $E_{XUV}(t)$ is the electric field of the XUV pulse, $D_k$ is the transition dipole moment between the ground state and the electron-hole state at the Bloch wavevector, $k$, and $H_{kk'}(t)$ is the electron-hole Hamiltonian. We employ the Wannier-Mott model[10,11], and the electron-hole Hamiltonian is given by

$$H_{kk'}(t) = \delta_{kk'}\left[E_g + \frac{1}{2\mu}\left(\hbar k + \frac{e}{c}A_{IR}(t)\right)^2\right] - V_{kk'} \qquad (3)$$

where $E_g$ is the direct gap of the matter, $\mu$ is the effective electron-hole mass, and $V_{kk'}$ is the Coulomb interaction, which we model with the one-dimensional soft Coulomb interaction. Note that, by diagonalizing $H_{kk'}(t = 0)$, one obtains exciton states as bound states of electrons and holes as shown in Fig. 1c.

By solving the Schrödinger equation, Eq. (2), one can compute the time-evolving wavefunction $|\Psi(t)\rangle$ with Eq. (1) and physical observables can be computed by making use of it. For example, the induced electric current density can be computed as $J_{XUV}(t) = \langle\Psi(t)|\hat{J}|\Psi(t)\rangle/\Omega$, where $\hat{J}$ is the current operator and $\Omega$ is the crystal volume. Furthermore, the linear susceptibility $\chi_{exc}(\omega)$ can be evaluated as

$$\chi_{exc}(\omega) = i\frac{\sigma_{exc}(\omega)}{\omega} = \frac{i}{\omega}\frac{\tilde{J}_{XUV}(\omega)}{\tilde{E}_{XUV}(\omega)}, \qquad (4)$$

where $\sigma_{exc}(\omega)$ is the optical conductivity evaluated as the ratio of the current and the electric field in the frequency domain. We model the dielectric function of MgF$_2$ by combining the core-exciton susceptibility $\chi_{exc}(\omega)$ and the valence contribution as

$$\epsilon_{MgF_2}(\omega) = \epsilon_{valence}(\omega) + 4\pi c\left[\chi_{exc}(\omega) + \frac{1}{3}\chi_{exc}(\omega + \Delta_{SO})\right], \qquad (5)$$

where $\epsilon_{valence}(\omega)$ is the valence contribution, $c$ is a fitting parameter, and $\Delta_{SO}$ is the spin-orbit split. By optimizing $\epsilon_{valence}(\omega)$ and $c$, the dielectric function and the reflectivity of MgF$_2$ can be well reproduced by the above model (see Fig. 1b).

With the final goal being to investigate the transient reflectivity of the XUV pulse under the presence of the IR pulse, we compute the electron dynamics by solving Eq. (2) with both the IR and XUV fields. Then, we further compute the current with the time-dependent wavefunction $|\Psi(t)\rangle$. Following the above procedure, Eq. (4) and Eq. (5), the transient dielectric function under the presence of the IR field and the corresponding transient reflectivity can be evaluated. The computed transient reflectivity by the Wannier-Mott model is shown in Fig. 1e.

In order to obtain further insight into the phenomena, we construct two idealized models. One is the *pure-exciton model*, and it is designed to exclude the atomic nature of the dynamics. The other is the *pure-crystal model*, and it is designed to exclude the crystalline nature. The pure exciton model is a three-level model constructed by the following three states; the ground state $|\Phi_{GS}\rangle$, the bright exciton state, and the dark exciton state (see Fig. 1c). Hence the pure-exciton model allows us to extract the nature of the discrete energy levels (atomic nature). Note that the pure exciton model is constructed by the subspace of the above Wannier-Mott model because the bright exciton state is the ground state of $H_{kk\prime}(t=0)$ and the dark exciton state is the first excited state. The pure crystal model, instead, consists of the conduction band (CB) but excludes the exciton states (see Fig. 1c). Such model can be realized by setting the electron-hole attraction $V_{kk\prime}$ to zero, and it is identical to the parabolic two-band model used to discuss the dynamical Franz-Keldysh effect [30]. Therefore, the pure crystal model allows us to extract the contribution purely from the intra-band motion (crystal nature) accelerated by the IR field. For a detailed discussion of the theoretical model, see Supplementary Information.

**Acknowledgments**

This project has received funding from the European Research Council (ERC) under the European Union's Horizon 2020 research and innovation programme (grant agreement No. 848411 title AuDACE). ML, GI and LP further acknowledge funding from MIUR PRIN aSTAR, Grant No. 2017RKWTMY. SS, HH, UDG and AR were supported by the European Research Council (ERC-2015-AdG-694097) and Grupos Consolidados UPV/EHU (IT1249-19).

**Author Contributions**

G.D.L, B.M and G.I performed the measurements. Together with R.B.V, M.N. and ML, they also evaluated and analysed the results. G.D.L, B.M, G.I, ML, F.F and L.P designed and built the reflectometer. S.A.S., H.H., U.D.G., and A.R. developed the theoretical models and performed the calculations. ML wrote the manuscript. All authors contributed to the analysis and interpretation of the experimental and theoretical results and writing of the manuscript.

**Competing interests**

The authors declare no competing interests